\documentclass[conference]{IEEEtran}
\IEEEoverridecommandlockouts
\usepackage{cite}
\usepackage[normalem]{ulem}
\usepackage{amsmath,amssymb,amsfonts}
\usepackage{algorithmic}
\usepackage{pgfplots, pgfplotstable}
\usepackage{graphicx}
\usepackage{pgfplots}
\usepackage{pgfplotstable}
\pgfplotsset{compat=1.16}
\usepackage{textcomp}
\usepackage{xcolor}
\usepackage{lineno,hyperref}
\usepackage{subfigure}
\usepackage{rotating}
\usepackage[ruled]{algorithm2e}
\usepackage{soul}
\usepackage{framed}
\usepackage{graphicx}
\usepackage{float}
\usepackage{url}
\usepackage{rotating}
\usepackage{fixltx2e}

\usepackage{verbatim}
\usepackage{multirow}

\usepackage{pgfplots, pgfplotstable}
\usepackage{scalefnt}
\usepackage{tikz}
\usetikzlibrary{shapes.gates.logic.US, trees, positioning, arrows}
\usetikzlibrary{shapes.multipart}
\usetikzlibrary{shapes,arrows}
\usepackage{xcolor}

\def\BibTeX{{\rm B\kern-.05em{\sc i\kern-.025em b}\kern-.08em
    T\kern-.1667em\lower.7ex\hbox{E}\kern-.125emX}}

\usepackage{xcolor}
\def\BibTeX{{\rm B\kern-.05em{\sc i\kern-.025em b}\kern-.08em
    T\kern-.1667em\lower.7ex\hbox{E}\kern-.125emX}}
    
\begin{document}

\title{Dependency Dilemmas: A Comparative Study of Independent and Dependent Artifacts in Maven Central Ecosystem\\
}
\author{
   Mehedi Hasan Shanto$^{\delta}$ , Muhammad Asaduzzaman, Manishankar Mondal$^{\delta}$, Shaiful Chowdhury$^{\dagger}$ \\
   \textit{$^{\delta}$Computer Science and Engineering Discipline, Khulna University, Bangladesh}\\
    \textit{School of Computer Science, University of Windsor, Canada} \\
    \textit{$^{\dagger}$Department of Computer Science, University of Manitoba, Canada} \\
    shanto1832@cseku.ac.bd$^{\delta}$, masaduzz@uwindsor.ca, 
    \\ mshankar@cse.ku.ac.bd$^{\delta}$, shaiful.chowdhury@umanitoba.ca$^{\dagger}$}

\vspace{-2mm}
\newcommand{\md}[1]{\textcolor{blue}{{\it [Muhammad says: #1]}}}
\maketitle

\begin{abstract}
Maven Central ecosystem forms the backbone of Java dependency management, hosting artifacts that vary significantly in their adoption, security, and ecosystem roles. Artifact reuse is fundamental in software development, and ecosystems like Maven facilitate this process. However, prior studies predominantly analyzed popular artifacts with numerous dependencies, leaving those without incoming dependencies (i.e., independent artifacts) unexplored. In this study, we analyzed 658,078 artifacts, of which 635,003 had at least one release. Among these, 93,101 artifacts (15.4\%) were identified as independent (in-degree = 0), while the rest were classified as dependent. We looked at the impact of individual artifacts using PageRank and out-degree centrality and discovered that independent artifacts were very important to the ecosystem. Further analysis using 18 different metrics revealed several advantages and comparability of independent artifacts with dependent artifacts: comparable popularity (25.58 vs. 7.30), fewer vulnerabilities (60 CVEs vs. 179 CVEs), and zero propagated vulnerabilities. These findings suggest that independent artifacts might be a beneficial choice for dependencies but have some maintenance issues. Therefore, developers should carefully incorporate independent artifacts into their projects, and artifact maintainers should prioritize this group of artifacts to mitigate the risk of transitive vulnerability propagation and improve software sustainability.
\end{abstract}


\section{Introduction}
Modern software development has made the usage of third-party artifacts a pillar, allowing teams to increase productivity, cut expenses, and shorten development times \cite{dev_productivity},\cite{productivity2}\cite{repoduction_artifacts}. Artifact repositories like Maven Central, npm, and PyPI are essential for enabling artifact sharing and code reuse, thereby promoting dynamic software ecosystems. Maven Central supports the Java ecosystem by providing millions of software components for smooth integration into software projects.

Despite the significant benefits of third-party artifact reuse, such as increased productivity and ecosystem collaboration, there are still some issues that require resolution. This includes maintenance burden, elevated susceptibility concerns, and possible legal ramifications \cite{risk1}\cite{mitigatingvul},\cite{popular1}. Previous research \cite{mitigatingvul} showed that upstream libraries that block fixes may cause vulnerabilities in dependent applications to stay open. A prior study on transitive vulnerabilities indicated that almost one-third of Maven packages in the dataset were deemed insecure when transitive dependencies were taken into account \cite{transitivity}. This highlights the hidden threat that transitive dependencies pose to software security.

 Prior studies have focused on popular artifacts with numerous incoming dependencies and transitive connections \cite{popular1}. But there is a lack of research on a specific type of artifact in these ecosystems: those without any incoming dependencies, referred to as ``independent artifacts.". One may think that these artifacts are trivial since they lack dependencies; nonetheless, this belief may underestimate their potential contributions to functionality, security, and ecosystem stability.

To address this gap, we question whether independent artifacts are as insignificant as they appear. Because they don't depend on other artifacts, independent artifacts can lower the risks that come with transitive vulnerabilities. Therefore, it is better to choose artifacts with no dependency while there are independent alternative artifacts available instead of dependent artifacts.  Additionally, the standalone nature of these artifacts indicates that developers have entirely authored them, potentially offering unique functionality and security advantages.
Chowdhury et al. investigated the trivial packages for the npm ecosystem~\cite{triiviality}. The term ``trivial packages" refers to small, lightweight npm packages that offer basic functionalities, sometimes with only a few lines of code, yet play an essential role in the ecosystem. Another study investigated the reasons behind using these trivial packages in npm~\cite{whydo}. They revealed that trivial packages are not trivial in reality. We think independent artifacts of Maven Central may play the same role as trivial packages in npm; therefore, we plan to investigate the PageRank and degree centrality to find out the importance of the artifacts.


This paper addresses the following three research questions:\\
\textbf{RQ1: How prevalent are independent artifacts in Maven ecosystem?}

In this study, we investigate independent artifacts and their potentiality. In the context of our study, the term ``potentiality" refers to the possible benefits and impact of independent artifacts in software development, particularly in terms of security and maintainability.  We identified 15.4\% of our studied artifacts as independent. This is a good number to investigate, as many other artifacts might depend on them. 

\textbf{RQ2: Ecosystem influence: Are independent artifacts influential in Maven Central?}

Since independent artifacts have no dependencies, one may think that they are useless or that they have no influence on the ecosystem. For that reason, we want to investigate the influence of independent artifacts in Maven Central ecosystem. Using PageRank\cite{pagerank} and degree centrality \cite{degree}, we evaluated the ecosystem influence of independent artifacts. To avoid bias in the degree centrality metric, we only consider the out\_degree for both independent and dependent artifacts.

\textbf{RQ3: Comparison: Are there any differences between independent and dependent artifacts?}

We want to investigate whether the two groups are significantly different from each other. One can argue that fixing vulnerabilities in independent artifacts is easier than in dependent artifacts because they do not depend on other artifacts. We compared independent and dependent artifacts across 18 different metrics from a usability and maintenance perspective, including update frequency, popularity, own vulnerability counts, and so on. 

The findings indicate several benefits of treating them as dependencies, despite their poor maintenance. The replication package of our study is available at \cite{online}\\


\begin{table}[tp]
	\caption{Distribution of artifacts in Maven Central}
	\label{Distribution}
	\begin{center}
	\footnotesize
		\renewcommand {\arraystretch}{1.2}
		\begin{tabular}{c|c }\hline
			
			Artifacts & Number	\\ \hline\hline 
                Total artifacts & 6,58,078\\
                Total artifacts with minimum 1 release & 6,35,003 \\
			Active artifacts & 2,89,721   \\
		  Active independent artifacts & 22,081 \\			
          	Active dependent artifacts & 2,67,640  \\

			\hline
			\hline
			
		\end{tabular}
	\end{center}
\end{table}
\vspace{-7mm}

\begin{figure}[t]

\pgfplotstableread{
0	7.9 
1	7.1 
2   8.0
3   6.8
4   4.2
5   6.4
6   4.8
7   4.7
8   7.2
9   4.1
10  6.0
11  4.3
12  4.9
}\dataset

\begin{tikzpicture}
\begin{axis}[
	ybar=0.25 cm,
	bar width=0.35 cm,
    width=0.52\textwidth,
    height=5 cm,
    ymin=0,
    ymax=10, 
    ylabel={(\%) of independent artifacts (active)},
    yticklabel style={font=\scriptsize},
    xtick=data,
    xticklabels = { 2011 , 2012 , 2013 , 2014 , 2015 , 2016 , 2017 , 2018 , 2019 , 2020 , 2021 , 2022, 2023 },
    xticklabel style={yshift=0ex, rotate=90, font=\scriptsize},
    major x tick style = {opacity=0},
    minor x tick num = 0,
    minor tick length = 1ex, 
    ymajorgrids = true,
    legend entries={(\%) of independent artifacts (active) to all published  artifacts (active) in Maven
},
    nodes near coords,                            
    nodes near coords style={font=\scriptsize, anchor=south, yshift=1mm}, 
    every node near coord/.append style={
        /pgf/number format/fixed,                 
        /pgf/number format/precision=1,           
    },
    legend style={
				at={(0.48,1.05)}, 
				anchor=south,
				legend columns=0,
				font=\scriptsize,
				text width=0.442\textwidth,
				minimum height=0.12in,
				},
]

\addplot[draw=black, fill=cyan!50] table[x index=0,y index=1] \dataset;

\end{axis}
\end{tikzpicture}

\vspace{-4mm}
\caption{Distribution of active independent artifacts in Maven Central over the years}
\label{rq1}

\end{figure}
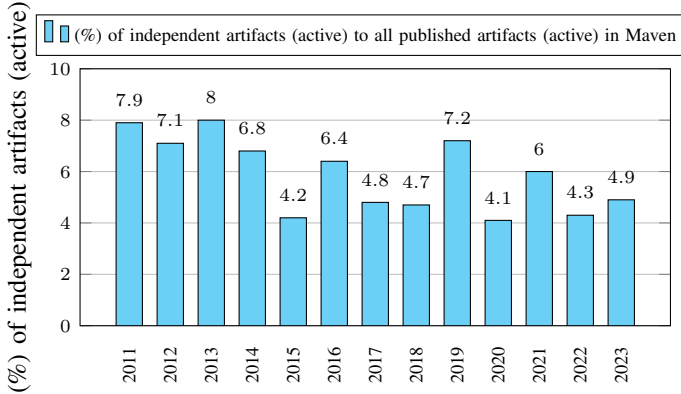

\section{Data Collection}
To conduct our study, we utilized the Goblin dataset \cite{gobline2}, a comprehensive resource containing metadata and dependency information for artifacts in Maven ecosystem. The dataset, released in its latest version, includes 0.65 million unique artifacts and 14.5 million releases. Each artifact represents a library or component within the ecosystem, with releases corresponding to different versions.

We focused on artifacts with at least one release to ensure relevance to the ecosystem. Out of 658,078 artifacts, 635,003 have at least one release. The discrepancy arises due to artifacts that exist in the repository but have not yet undergone a formal release process. These may include placeholders, unpublished versions, or artifacts that were never fully deployed.These artifacts were further categorized into two groups based on their incoming dependencies:

\begin{itemize}
    \item \textbf{Independent artifacts:} These artifacts do not hold any dependencies with the upstream artifacts (number of incoming dependencies = 0).

    \item \textbf{Dependent artifacts:} These are defined as artifacts that have at least one incoming dependency (number of incoming dependency \verb|>| 0).
\end{itemize}

We further refined these groups by identifying \textbf{active artifacts} that have dependent artifacts within the ecosystem. We extracted crucial details for each artifact to perform a more comprehensive analysis. Vulnerability information, derived from CVE records, was collected to calculate the number of artifacts containing vulnerabilities and the severity of vulnerabilities (e.g., critical, high) characterized by \cite{cvss}. 

\textbf{Features Extraction: }
We leveraged Cypher queries \cite{cipher} to retrieve relevant metadata from Maven Central Neo4j dataset provided by \cite{gobline2}, including releases, release frequency, dependencies, usage, popularity trends, and vulnerabilities.

For our quantitative analysis, we look at the number of active independent and dependent artifacts, their number of releases, the number of dependencies and dependents of each artifact's most recent release, the frequency and delay of releases, how often they are used, and the number of artifacts that are linked to common vulnerability exposures (CVE), common weakness enumeration (CWE), and propagated vulnerabilities. CWE is a standard for classifying and describing the types of weaknesses that can lead to vulnerabilities \cite{cwe}.

\textbf{CVE Information Extraction: }
We extracted the CVE information for releases of each artifact. We also extracted the CWE, CVE name, and CVE severity information from CVE metadata. We found 4 types of CVE severity, including critical, high, moderate, and low.  

\textbf{GiitHub Repository Analysis:} We want to access the information about their repository structures to identify how healthy the independent and dependent artifacts are. We retrieved artifact identifiers and the most recent release versions from Maven Central Neo4j dataset (Version 2024-08-30).
Based on the data, we generated Maven URLs to access pom.xml files of packages. We collected GitHub repository URLs mentioned in the pom.xml files and found 74,435 GitHub repositories. Upon filtering and mapping our collected artifacts with those repositories, we identified a total of 43,529 active artifacts: 41,153 dependent and 2,376 independent artifacts hosted on GitHub. We extracted various metrics from their repositories, such as the number of subscribers, the number of open issues, the number of closed issues, the number of pull requests, and the number of contributors. We also evaluated how many of the mapped repositories have README.md files, the number of repositories with missing licenses, and their used programming languages. We used this information to understand their maintenance.  

\begin{table*}[tp]
\caption{Studied features}
\label{tab:pvalue_cohen}
\centering
\footnotesize
\renewcommand{\arraystretch}{1.2}
\begin{tabular}{|c|c|c|c|}
\hline
 \textbf{Feature} & \textbf{Mean/Median/Min/Max (Dependent)} & \textbf{Mean/Median/Min/Max (Independent)} \\ \hline \hline
\multicolumn{3}{|c|}{\textbf{Dataset Analysis}}\\
\hline
PageRank & 11.70/0.44/0.15/333563 & 26.03/0.40/0.15/100950   \\ \hline
(Out) Degree Centrality & 1.4E-4/5.06E-6/3.6E-7/1.00 & 3.53E-4/5.04E-6/6.3E-7/1.00  \\ \hline
NumReleases & 31.0/8.0/1.0/3281.0 & 19.0/3.0/1.0/3614.0  \\ \hline
NumDependent & 391.0/14.0/1.0/2.7E+6 & 557.0/8.0/1.0/1.57E+6 \\ \hline
Popularity (Past Yr.) & 7.30/0.0/0.0/112049.0 & 25.8/0.0/0.0/128031.0  \\ \hline

ReleaseFrequency & 0.077/0.02/0.0/55.0 & 0.057/0.006/0.0/9.0  \\ \hline
NumCVE (Last Release) & 179.0 & 60.0  \\ \hline
NumAtrtifactsWithCVE & 793.0 & 69.0  \\ \hline
NumCWE(Last Release) & 453.0 & 318.0  \\ \hline
\hline \hline

\multicolumn{3}{|c|}{\textbf{GitHub Repository Analysis} (Investigated 43529 Repositories, Dependent: 41153, Independent: 2376)}\\ \hline
 NumSubscribers & 40.0/7.0/0.0/6616.0 & 47.0/6.0/0.0/3325.0  \\ 
 \hline
NumOpenIssue & 12.0/8.0/0.0/30.0 & 8.5/2.0/0.0/30.0  \\ \hline

NumClosedIssue & 22.2/30.0/0.0/30.0 & 17.0/24.0/0.0/30.0  \\
\hline
NumPullRequests & 22.08/8.0/0.0/30.0 & 23.9/21.0/0.0/87.0  \\
\hline
NumContributors & 13.5/7.0/0.0/30.0 & 9.4/4.0/0.0/30.0  \\ \hline
NumCommits & 28.0/30.0/0.0/30.0 & 28.0/30.0/0.0/30.0  \\
\hline
ReadmeAvailable & (37457/41153) = 91.0\% & (2037/2376) = 85.0\%  \\
\hline
MissingLicences & (3354/41153) = 8.1\%& (310/2376) = 13.0\%  \\
\hline
Top3UsedLanguages & Scala:15376, Kotline: 12074, Java: 11016 & Java:1342, Js: 384, Kotlin: 226  \\
\hline
\end{tabular}
\end{table*}

Our extracted features and results are listed in Table \ref{tab:pvalue_cohen}.
Using these metrics, we answered our research questions.
\vspace{-2mm}
\section{Findings}
\subsection{Prevalence in Ecosystem}
\textbf{RQ1: How prevalent are independent artifacts in Maven Central ecosystem?}\\
Independent artifacts in Maven Central hold significant potential, with 22,081 active artifacts (3.4\% of total artifacts and 7.6\% of all active artifacts) identified as independent (Table \ref{Distribution}). The identification of independent artifacts was based on their in-degree value in the dependency graph, where artifacts with an in-degree of zero were classified as independent. This classification was derived from the Goblin \cite{gobline2} dataset, which provides Maven Central dependency relationships. These artifacts often have fewer external vulnerabilities, simplified maintenance, and reduced compatibility concerns compared to their dependent counterparts. The low prevalence of independent artifacts suggests the ecosystem is heavily interdependent, reflecting the modular and reusable nature of software development. Independent artifacts are useful when we need lightweight, self-contained, or stand-alone solutions. 

The bar plot (see Figure \ref{rq1}) shows fluctuations in the annual percentage of independent artifacts published in Maven Central. Some years have higher proportions of independent artifacts, while others, like 2023, show a decline. This indicates a shift in artifact publication with dependent artifacts dominating. Independent artifacts can serve as foundational libraries for complex systems, fostering trust and adaptability in software development. Therefore, identifying and nurturing high-quality independent artifacts is crucial in Maven Central. 

\subsection{Influence of independent artifacts in the ecosystem}
\textbf{RQ2: Ecosystem influence: Are independent artifacts influential in Maven Central ecosystem?}

PageRank is a link analysis algorithm that assigns a numerical weighting to artifacts based on their connectivity within the dependency network. From the analysis of \textit{PageRank} \cite{pagerank} and \textit{(Out) Degree Centrality} \cite{degree}, it is evident that dependent artifacts are more influential in Maven Central ecosystem than independent artifacts. The median PageRank value for dependent artifacts stands at 0.44, which is slightly higher than 0.40 for independent artifacts. Although the difference is not significant, it highlights the stronger centrality of dependent artifacts within the network. Furthermore, the maximum PageRank value for dependent artifacts reaches 333,563, significantly surpassing the maximum value of 100,950 for independent artifacts. Similarly, the median \textit{(Out) degree centrality} value for dependent artifacts (5.06E-6) is notably larger than that of independent artifacts (6.3E-7), reflecting stronger connectivity and greater influence in the dependency network. It is clear that artifacts that depend on other nodes in the network interact with more of them, which makes their effects on the ecosystem stronger.

Despite the fact that independent artifacts fall behind in PageRank and degree centrality, it is important to acknowledge their presence. The differences in median values, though statistically significant, remain relatively small. This indicates that independent artifacts remain relevant within the ecosystem, especially when considering their unique position. Since independent artifacts do not rely on other components, they can serve as stable alternatives that do not propagate vulnerabilities downstream. Their lack of dependencies makes them a safer option in specific contexts, even though their influence and connectivity in the ecosystem are comparatively lower.

\subsection{Comparison between independent and dependent artifacts?}
\textbf{RQ3: Are there any differences between independent and dependent groups of artifacts?}

Dependent and independent artifacts exhibit notable differences across usability and maintenance metrics with critical implications for their use and maintenance. Dependent artifacts have more releases (31 on average vs. 19) and more dependent artifacts (median 14 vs. 8), which shows how widely they are used and how well they fit into Maven Central ecosystem. However, the significantly higher numbers of CVEs (793 vs. 69) and CWEs (453 vs. 318) indicate that these artifacts face greater security risks. Some major worries are brought to light by this: flaws in dependent artifacts could spread to other parts, creating problems throughout the dependency network.

On the other hand, independent artifacts, while not contributing to dependency-related vulnerabilities, show poor performance in terms of release frequency (median 0.006 vs. 0.02 for dependent artifacts) and maintainability metrics. Although it may seem that independent artifacts are easier to maintain since they lack dependencies, the data reveals otherwise. They exhibit lower subscriber counts (median 6 vs. 7) and fewer contributors (median 4 vs. 7), which suggests smaller development teams and reduced maintenance capacity. Moreover, while independent artifacts show relatively higher pull request activity (median 21 vs. 8), this effort does not translate into frequent updates or higher overall quality. The documentation and licensing gaps further exacerbate their maintainability issues, as 85\% of independent artifacts provide README files compared to 91\% for dependent artifacts, and 13\% lack licenses vs. 8.1\% for dependent artifacts. 

These findings underscore the importance of exercising caution when selecting independent artifacts. Because they do not depend on other artifacts, vulnerabilities are less likely to spread. However, due to their less frequent updates and lower maintainability metrics, they may stagnate, leaving potential vulnerabilities unaddressed for extended periods. The median value of the release frequency metric indicates that the average time taken to publish a new release for independent artifacts is 166 days, whereas for dependent artifacts the value is 50 days. Maintainers of independent artifacts should prioritize regular updates, improve documentation, and enhance community involvement to unlock their potential within Maven ecosystem. Users also should carefully assess the trade-offs between stability and maintainability when considering independent artifacts as alternatives to dependent ones.

\section{Discussion }
We investigated the Goblin \cite{gobline2} dataset, which provides useful information for analyzing the dependency of artifacts from Maven Central. We designed and addressed three research questions, with a particular focus on understanding the prevalence, importance, and maintenance of independent artifacts within the ecosystem. Initially, we measured their evolution and prevalence, concluding that their number is not insignificant. They are not vulnerable to upstream dependencies but they can propagate their own vulnerabilities. We also found that a large number of artifacts depend on them. Thus, they can be seen as an alternative to the dependent artifacts. 
They are not as well-structured as the dependent artifacts regarding release frequency, the number of contributors, the absence of licenses, and the scarcity of README files for developers. The broader implications of independent artifacts are mitigating transitive vulnerabilities and offering self-contained alternatives for dependency management. Practical applications include their potential use in security-sensitive environments where dependency minimization is crucial and their adoption in long-term maintenance strategies to reduce reliance on frequently changing third-party artifacts. Future research directions include developer surveys to assess real-world adoption challenges, automated tools to enhance the maintainability of independent artifacts, and further analysis of their impact on software supply chain security.

\section{Conclusion}
We conducted an empirical analysis to differentiate independent and dependent artifacts in Maven Central ecosystem. We found that independent artifacts can play an important role in Maven Central ecosystem while having no dependencies. However, they can suffer from the vulnerabilities of their own. As the release frequencies are comparatively low, fixing vulnerabilities may take time. Further analysis of GitHub repositories indicates a lower level of maintenance for independent artifacts than for dependent artifacts. Based on the above findings, we can conclude our study by providing insightful suggestions to both artifact maintainers and the users of those artifacts. Users should exercise caution when considering them as alternatives to dependent artifacts, and maintainers should prioritize this group of artifacts in terms of regular updates, fixing vulnerabilities, and immediate responses to issue reports. In the future, we will survey developers to understand the practical challenges and benefits of using independent artifacts.

\bibliography{References}

\end{document}